# Growth of large area graphene from sputtered films

*Genhua Pan[1]\*, Mark Heath[1,2], David Horsell[2] and M. Lesley Wears[2]*

[1] Wolfson Nanomaterials & Devices Laboratory, Faculty of Science and Technology, University of Plymouth, Devon, PL4 8AA, UK

[2] College of Engineering, Mathematics and Physical Sciences, University of Exeter, Exeter, EX4 4QF, UK

**ABSTRACT**: Techniques for mass-production of large area graphene using an industrial scale thin film deposition tool could be the key to the practical realization of a wide range of technological applications of this material. Here, we demonstrate the growth of large area polycrystalline graphene from sputtered films (a carbon-containing layer and a metallic layer) using in-situ or ex-situ rapid thermal processing in the temperature range from 650 to 1000 $^{o}$C. It was found that graphene always grows on the top surface of the stack, in close contact with the Ni or Ni-silicide. Raman spectra typical of high quality exfoliated monolayer graphene were obtained for samples under optimised conditions. A fast cooling rate was found to be essential to the formation of monolayer graphene. Samples with Ni atop SiC produced the best monolayer graphene spectra with ~40% surface area coverage, whereas samples with Ni below SiC produced poorer quality graphene but 99% coverage. The flexibility of the sputtering process allows further optimization of the growth, with possibility of transferring the graphene to any insulator substrate in vacuum. We present a potential route for the production of graphene-on-insulator wafers, which would facilitate easy integration of graphene into modern semiconductor device process flows.

\* Corresponding author: Tel: +441752586258. Email: gpan@plymouth.ac.uk

## 1. Introduction

Graphene has attracted considerable interest, particularly because of its wide range of potential applications [1-6]. To take many of these applications to an industrial level requires large scale growth of high quality graphene on device compatible substrates. To date, this has mainly been achieved via chemical vapour deposition (CVD) [7] and single crystal SiC epitaxial [8] growth routes. CVD graphene has been synthesised on various metal substrates such as ruthenium [9], iridium [10], platinum[11], nickel [12, 13] and copper [7], and even in the absence of a substrate [14]. Though suitable for mass production[15], the need to transfer these films to different substrates has so far constrained its up-scaling to roll-to-roll production methods. Epitaxially grown graphene has been demonstrated to be a viable route to the production of electronic devices, such as field effect transistors[16]; however, SiC wafers are expensive and, unless SiC is required in the device, again the graphene needs to be transferred. Graphene has also been synthesised via the SiC route by using rapid thermal processing (RTP) at a temperature of 1100 $^{o}$C [17]. Lower temperature synthesis of few-layer graphene has also been reported [18] by deposition of a Ni layer atop single crystal SiC substrate with RTP temperatures around 750 $^{o}$C.

Here we show the growth of large area graphene from sputtered SiC films either atop or underneath a Ni layer. We also demonstrate the feasibility of sputtering the final substrate material on top of the graphene within the same process cycle, thereby avoiding the need to transfer graphene in a subsequent processing step.

## 2. Experimental

The deposition of films was carried out in a three-target RF diode and magnetron sputtering machine. Typical base pressure was $2 \times 10^{-7}$ Torr and Ar pressure for deposition, 3 mTorr. All targets are 6 inch in diameter and have a typical purity of 99.99%. No substrate heating was applied during the deposition of all the layers. The substrates used in the work were single side polished Si wafers with a 300 nm-thick thermally oxidized $SiO_2$ layer, which was referred to as the 'Si wafer' throughout the paper.

Samples were subjected to RTP either in-situ or ex-situ. The in-situ RTP was carried out in the same vacuum chamber as the sputter-deposition using a 500W halogen lamp and the sample was heated for 10 minutes to reach a maximum temperature of 650 $^o$C before cooling down naturally by switching off the halogen lamp (See Fig.S1 for heating and cooling curves.) The in-situ RTP used in the work had a heating and cooling rate that was restricted by the maximum heating power available and large thermal mass of the sample holder. The best graphene films were grown using ex-situ RTP.

The ex-situ RTP was carried out in a commercial RTP apparatus with a quartz chamber. The RTP apparatus provides a much faster heating and cooling rate than the in-situ system (see Fig.S1). There are eleven 1.5 kW halogen lamps available for rapid temperature ramp up at a typical rate of 100 $^o$C/s up to 1200 $^o$C. It also has a rapid cooling capability due to a cold chamber wall design and a very small thermal mass of the quartz sample holder. The sample chamber was purged with Ar gas for one hour before the RTP process, which was also carried out in Ar atmosphere. We found this is necessary to prevent the potential oxidation of the sample. In order to achieve the growth of good quality graphene, samples need to be annealed fresh, particularly for samples with Ni on top: surface oxidation has an adverse effect on the formation of graphene.

As-grown and transferred graphene films were examined by Raman spectroscopy (with a 532 nm laser). Raman maps across large areas were carried out with a step size of 30 μm. Film thicknesses and surface morphology were characterised by AFM. Bragg-Brentano geometry and grazing-incidence x-ray diffraction at 1° angle of incidence were used for the microstructural and crystallographic characterisation of the metal and SiC films.

For device fabrication, a rectangle of graphene film of 18μm by 5μm was defined by e-beam lithography and a reactive ion-etching step. Cr/Au (5/70 nm) electrical contacts were defined by a second lithographic step and deposited using a thermal evaporator. The measurement of the devices was carried out under vacuum ($5\times10^{-4}$ mbar) using standard low-frequency lock-in techniques. The source-drain current was fixed at 100 nA.

## 3. Results and Discussion

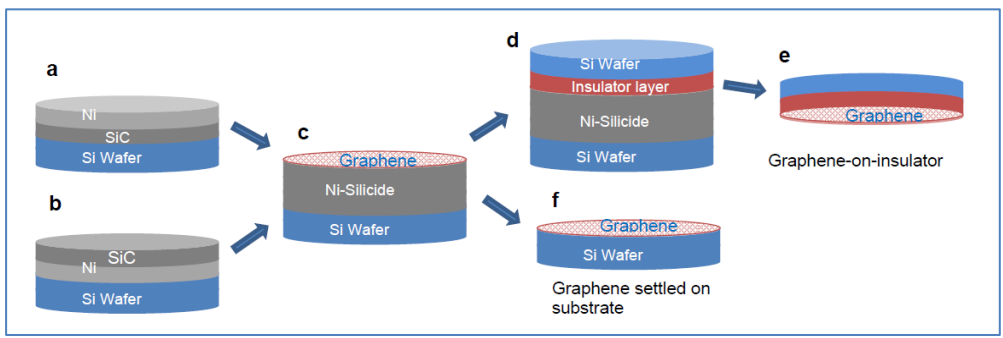

**Figure 1 - Schematic illustration of the graphene growth process.** Sputter deposition of a carbon containing layer (SiC) and a metal film (Ni) on a Si wafer in the order of either SiC/Ni **a**, or Ni/SiC **b**. **c,** Rapid thermal processing step, resulting in Ni-silicidation and the formation of graphene atop the Ni-silicides upon cooling. **d**, An insulating layer sputtered directly onto the graphene and subsequently bound to a silicon wafer. **e**, Removal of the sacrificial growth wafer. **f**, An alternative step where graphene is transferred to the silicon wafer by dissolving the Ni-Silicide.

Figure 1 shows our graphene growth process. A carbon-containing film (SiC) and a metal film (Ni) were deposited by sputtering on to a Si substrate in the order of either substrate/SiC/Ni, Fig.1a, or substrate/Ni/SiC, Fig.1b. (Here we consider only SiC and Ni.

However, we have also found that amorphous carbon and Pt can be used instead of SiC and Ni: some initial results from C/Ni and SiC/Pt stacks are shown in Fig.S3 of Supplementary Information.) After deposition, the stacks were rapid-thermal-processed either in-situ or ex-situ. (Section 1 of Supplementary Information provides further details.) We found that graphene always grows on the top surface of the stack, irrespective of the deposition sequence of the two layers, Fig.1c. After RTP, two possible routes for transferring the graphene to insulating substrates were used, as shown in Fig.1d,e and Fig.1f, respectively. In Fig.1f, the Ni-silicide is dissolved in an HCl solution and the graphene settles on the Si substrate as a loosely adhered film, which is an approach already employed [12],[17],[18]. In Fig.1d and e, an SiO$_2$ layer serving as a new substrate for the graphene is deposited on to the as-grown graphene in the same evacuation by sputtering after the RTP. A supporting Si substrate can then be directly bonded to the new insulator, Fig.1d. The final wafer is obtained, after the removal of the original substrate by a Si etcher and an acid solution (to remove the Ni silicide), Fig.1e. This vacuum deposition of the new substrate ensures a good adhesion between it and the graphene film.

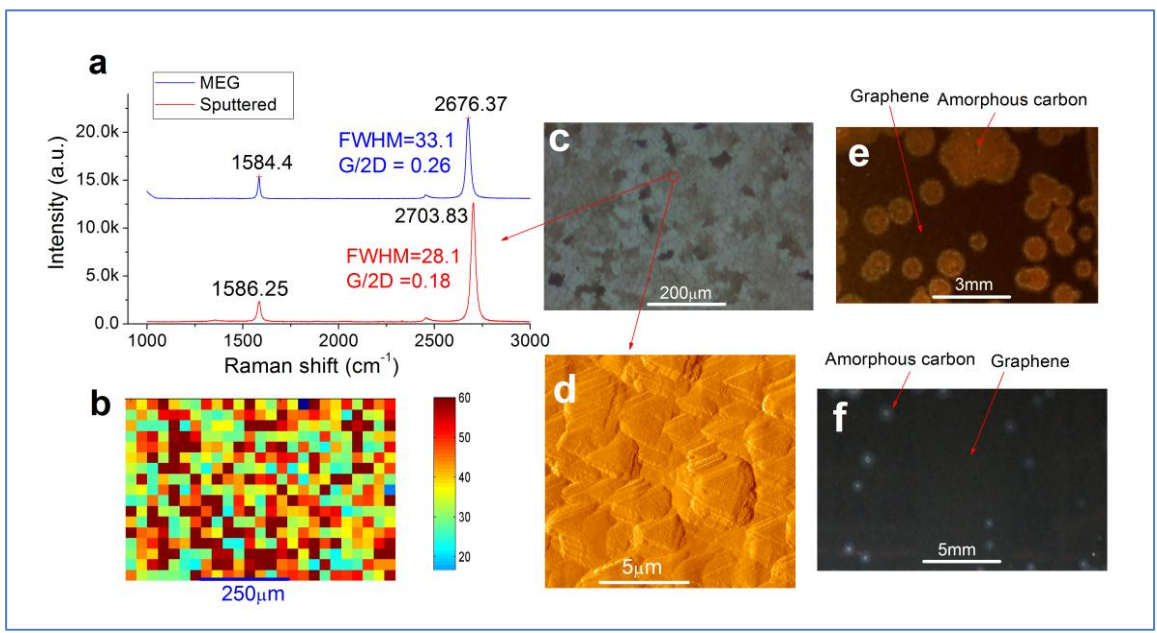

**Figure 2 - Typical characteristics of as-grown graphene.** **a**, Raman spectrum of as-grown graphene (red) of S1(sub/SiC(50nm)/Ni500(nm)). A spectrum of mechanically exfoliated monolayer graphene (MEG) on Si/SiO$_2$ substrate measured by the same Raman spectrometer is also shown for comparison (blue). Both spectra were normalised with the same G peak height. **b**, Map of the FWHM of the 2D band for a sample area of 750µmx500µm. **c**, Optical micrograph of the sample surface, where graphene appears as dark-gray in colour. **d,** AFM image of the polycrystalline graphene surface showing typical grain sizes of 1.5-2.5µm. **e and f**, Camera shots showing distinctive areas with and without graphene on S3 (sub/SiC(50nm)/Ni(500nm) (e) and S4 with reversed layer sequence of sub/Ni500(nm)/SiC(50nm) (f).

Figure 2a shows a typical Raman spectrum of graphene from sample S1 which was grown on substrate/SiC(50nm)/Ni(500nm) with ex-situ RTP at 1000 °C for 2 minutes. For comparison, a spectrum of mechanically exfoliated monolayer graphene obtained by the same spectrometer is also shown. The characteristic G and 2D bands can be clearly observed. The value of the full width at half maximum (FWHM) of the symmetric 2D band and the intensity ratio of the G to 2D peaks can be used to determine the number of layers [19-21], with typical values <35 cm$^{-1}$ and <0.5, respectively, for monolayer graphene. As can be seen, both the FWHM and the G/2D ratio for sample S1 indicate a monolayer. These values are smaller than that of epitaxial graphene [20] and are comparable with that of our comparative exfoliated sample. The lack of D band at 1350 cm$^{-1}$ and the almost identical G peak position of the two spectra suggest that the monolayer graphene has few defects or large grains [19]. Figure 2b shows a map of the FWHM for an area of 750µmx500µm mm of S1. The estimated monolayer surface coverage of the sample is around 40% for the mapped area. An optical micrograph of the as-grown sample is shown in Fig.2c, where the monolayer graphene is visible as the dark-gray coloured areas. The surface morphology can be seen from the AFM image of Fig.2d, showing that the film has a typical grain size of ~2µm. The areas of graphene of the samples are visually identifiable, as shown by the camera shots in Fig.2e and f for two samples with the same layer thicknesses (SiC(50nm) and Ni(500nm)) but reversed layer sequence (sub/SiC/Ni for S3 and sub/Ni/SiC for S4): the dark areas are graphene whilst the lighter areas are amorphous carbon. As can be seen, samples with SiC atop Ni, Fig.2f, produced much better graphene surface coverage than the samples with SiC underneath the Ni, Fig.2e, reaching nearly 99% for the sample size shown in Fig.2f. X-ray diffraction examination of the dark and lighter regions showed no significant difference in terms of phase structure and crystal texture (see Fig.S2 for details). This will be further discussed below.

To determine the optimal processing parameters, graphene was grown from stacks with various layer thicknesses, deposition sequences and RTP conditions. Typical results are shown in Fig. 3. Figure 3a shows two Raman spectra of sample S2 (sub/SiC(50nm)/Ni(200nm)) processed with in-situ RTP (red) and ex-situ RTP (blue), respectively. (Key differences and typical heating cooling curves of the two different RTP systems are given in section 1 and Fig.S1 of Supplementary Information.) The spectrum for the graphene by in-situ RTP has a FWHM value of the 2D band of 74.1 cm$^{-1}$ and G/2D intensity ratio of 1.77 indicating several-layer graphene [19]. The D/G intensity ratio of 0.17 also suggests that the graphene contains considerable nano-grains [22]. The AFM images for the sample, Fig.3a.1 and (a.2), confirm that the re-crystallisation process of the sample was incomplete with a large proportion of the area containing nano-grains with typical sizes of 20 to 30 nm, Fig.3a.2. In contrast, the same stack processed by ex-situ RTP (peak intensity of 68% for 2 minutes, which gave maximum temperature of 1000 °C; see curve Exsitu68pc in Fig.S1) produced graphene with a Raman 2D FWHM value of 47 cm$^{-1}$ and G/2D intensity ratio of 0.96, indicating a bi-layer. The AFM image in Fig.3a.3 shows that the sample is well crystallised with typical grain sizes in the range of 1 - 2.5 μm.

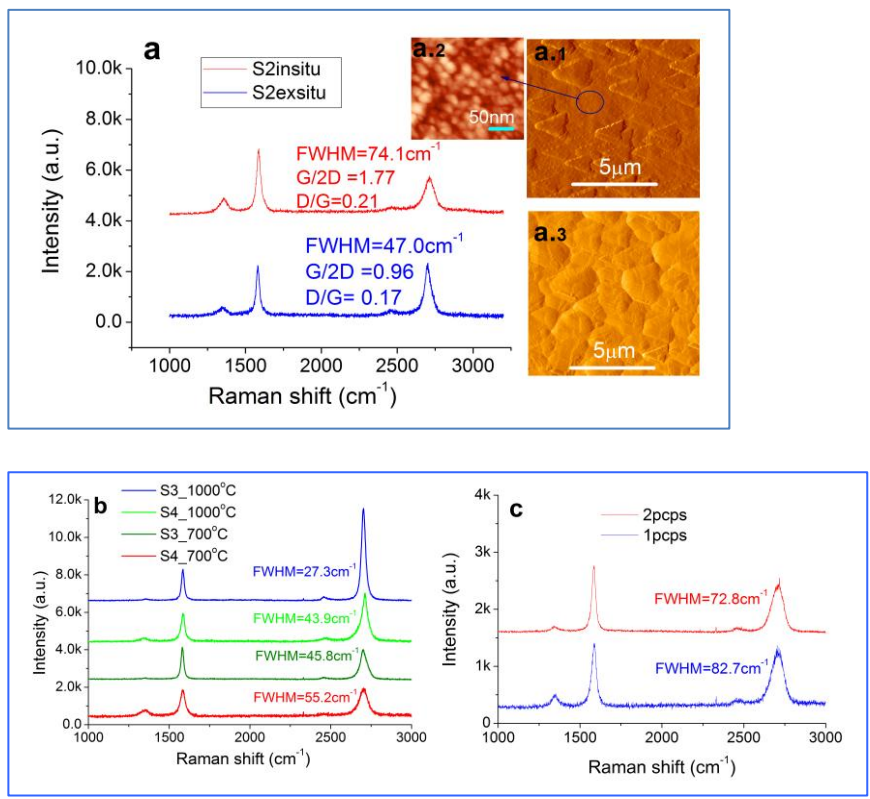

**Figure 3 - Raman spectra of as-grown graphene from samples with different Ni layer thicknesses, deposition sequences and RTP systems/conditions.** **a**, Raman spectra of S2 (SiC50nm/Ni200nm) with in-situ (red) and ex-situ (blue) RTP. AFM images for S2 in-situ are given in (**a.1**) and (**a.2**), and S2 ex-situ in (**a.3**). **b**, Raman spectra of samples S3 and S4 (with the same SiC and Ni thicknesses but reversed deposition sequence) processed with two different ex-situ RTP temperatures of 1000 °C (top 2) and 700 °C (bottom 2), respectively. **c**, Raman spectra of S4 with slower cooling rates.

Figure 3b shows the Raman spectra for two samples (S3 and S4), with the same SiC and Ni layer thicknesses but different deposition sequence and ex-situ RTP conditions. The top two spectra are for S3 and S4 processed with RTP at 1000 °C for 2 minutes (the same as for the ex-situ RTP for S2 in Fig. 3a). As can be seen, a spectrum indicative of good quality monolayer graphene was obtained for S3. The spectrum of S4 shows increased FWHM and G/2D ratio. However, the surface coverage of graphene for S3 and S4 are different with S4 showing much better coverage than S3, as already shown by the camera shots in Fig.2f and e, respectively. It is also worth noting from the Raman spectra of S2 and S3 prepared in the same ex-situ RTP conditions that thicker Ni films (500nm) favour the growth of higher quality monolayer graphene. The two spectra at the bottom of Fig.3b are for S3 and S4 processed with ex-situ RTP at a lower annealing temperature of 700 $^{o}$C for 2 minutes (refer to curve Exsitu10pc in Fig.S1). As can be seen, the FWHM values of the two spectra are 45.8 cm$^{-1}$ and 55.2 cm$^{-1}$, respectively, suggesting bi- or tri-layer graphene. Again, S3 (Ni atop SiC) exhibits slightly better Raman characteristics with smaller FWHM values and nearly negligible D band intensity.

Figure 3c shows two Raman spectra of S4 prepared under the same RTP condition of 1000$^{o}$C for 2 minutes but with slower cooling rate at 2%/s and 1%/s for the top and bottom spectrum, respectively (see Fig.S1 for cooling profile). These samples exhibit a much wider 2D band and also higher D band intensity than those with the slower cooling rate. In general, it was found that the heating rate was not crucial to the growth of graphene, but a faster cooling rate favours the growth of fewer layer graphene, which agrees with the results by Yu *et al*. [12].

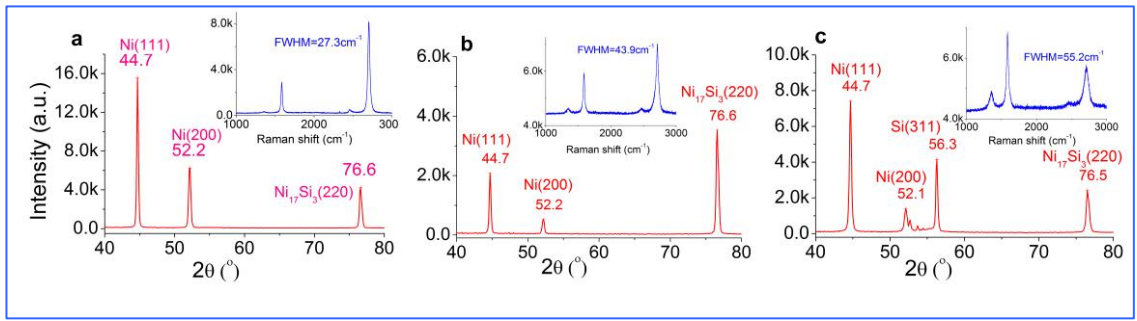

**Figure 4 - Grazing-incidence X-ray diffraction patterns for three samples.** Panel **a** is for S3; panel **b** for S4. Both samples were prepared with the same ex-situ RTP condition at 1000 $^{o}$C. Panel **c** is for S2 with in-situ RTP at 650 $^{o}$C. The inset in each panel is the corresponding Raman spectrum of the samples.

The crystal structure and crystallographic orientation of the films were examined by small angle grazing-incidence diffraction (GID) of x-rays. (Section 2 and Fig.S2 of the Supplementary Information give more details about the GID measurements.) Figure 4 shows the GID patterns of samples S3, S4 and S2. All three samples exhibit preferred Ni(111) and Ni$_{17}$Si$_3$(220) orientations, indicating good crystallites and complete Ni-silicidation. The contrast difference in peak intensities of the Ni(111), Ni(200) and Ni$_{17}$Si$_3$(220) of S3 and S4 suggests that there is a difference in the composition of the Ni-silicide across the depth of the films: the top surface of S3 is richer in Ni, and that of S4 is richer in Ni-silicide. (The structure on the top surface contributes more to the intensities of the diffraction peaks in GID.) This is natural, considering the different layer deposition sequence of the two samples and the relatively short RTP time (2 minutes), which was insufficient for a complete diffusion of the SiC throughout the whole 500 nm depth of the Ni film. It is also natural to expect that the carbon concentration across the film depth is also non-uniform. The better surface coverage of graphene in S4 compared to S3 is a result of a surface that is much richer in carbon. The richer carbon concentration may also contribute to the formation of more layers of graphene during the growth. The poorer concentration of carbon on the surface of S3 favours the growth of monolayers but results in a lower surface coverage. The reasons for the formation of amorphous carbon areas remain unclear: for instance, the existence of the Ni$_{17}$Si$_3$(220) phase does not appear to have any adverse effect on the growth of graphene. The major difference in the GID pattern of S2, Fig. 4c, is the presence of an additional peak of Si(311) and a relatively weaker and messy Ni(200) band. This suggests that the recrystallization and Ni silicidation processes were incomplete due to the lower RTP temperature (in agreement with the AFM images in Fig. 3a).

Two approaches were used for the transfer of graphene onto different insulating substrates, Fig. 1d-f. Figure 5a shows a typical Raman spectrum of graphene transferred to the silicon wafer that supported the stack, after the removal of the Ni and Ni-silicide layers in HCl. In comparison with the Raman spectrum of the as-grown graphene of the same sample S4 (Fig.3b), there is almost no change in the relative intensities of G and 2D peak and the value of the FWHM of the 2D band. However, the D peak intensity of the transferred graphene is slightly higher, which is most likely due to damage caused by the handling of the graphene film during the transfer process. (In general, the graphene film is robust and can be picked up by a piece of wafer without being visibly broken. In contrast, the amorphous carbon films were very fragile after removal of the metal.) On the microscopic scale, the transferred graphene is not smooth but appears to replicate the morphology of the Ni surface on which it grew. Figure 5a.2 is a topographic AFM image of the transferred graphene from S4, which shows well defined grains with typical sizes in the range of 1-2 μm.

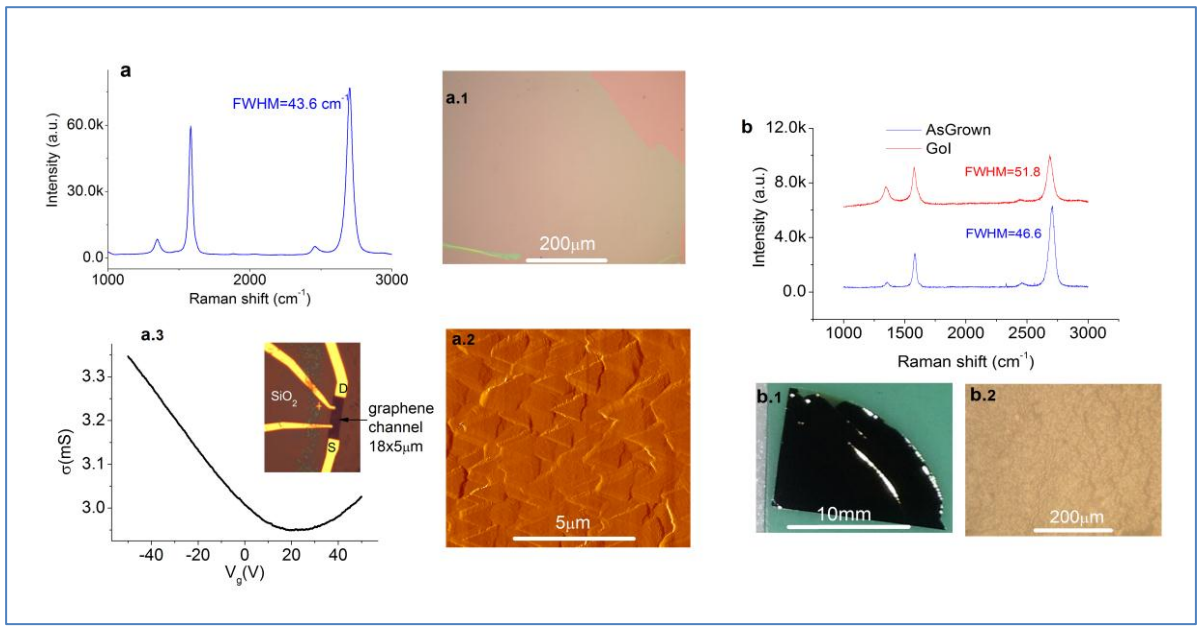

**Figure 5 - Characterisation of transferred graphene and GoI.** **a,** Raman spectrum of graphene settled on a Si substrate after removal of Ni-silicide in HCl: **a.1** and **a.2** show an optical micrograph and AFM image of the sample, respectively. **a.3** shows the electrical characteristics of a back-gated field effect transistor made from the transferred graphene from S4. The inset of **a.3** is an optical microscope picture of the device. **b**, Raman spectrum of graphene on the sputtered $SiO_2/Ti/SiO_2$ substrate (red) and before growth of the substrate (blue). The GoI is supported on heat-release tape and peeled off from the original substrate. **b.1** Image of the whole peeled-off sample with Ni-silicide/graphene/$SiO_2$(200nm) /Ti(200nm)/ $SiO_2$(200nm) layer stack on heat release tape. **b.2** Optical micrograph of the GoI after removal of the Ni-silicide in HCl.

To understand the electrical characteristics of the transferred films, we created a field-effect transistor from the bi/few-layer graphene transferred from S4, Fig. 5a.3. (The Si wafer in a.1 has a 300nm $SiO_2$ layer on its surface.) The gate modulation caused a change in resistivity of ~40 Ω (25 mS in conductivity) over the measurement range, with a charge neutrality point at +22 V. The gate modulation is not as effective as that expected from bi-layer exfoliated samples[23], but comparable - with graphene grown by other routes [17], [24]. We would expect this, because of the uneven structure of the film leading to a poor interface with the substrate and scattering from defects in the crystal caused by the growth and transfer.

To demonstrate the feasibility of producing graphene-on-insulator (GoI) by sputtering, Fig.1d,e, new insulating layers of $SiO_2$(200nm)/Ti(200nm)/$SiO_2$(200nm) were deposited on to the graphene after RTP. (See section 4 of Supplementary Information for details). The deposition of the $SiO_2$ was carried out using a magnetron target, with the initial layers of $SiO_2$ sputtered at power as low as 50 W to minimise the potential damage of plasma and high energy sputtering particles to the graphene [25-27]. A piece of heat-release tape was then applied to the top of the wafer. The stack could then peeled off the original substrate, as a result of the relatively poor adhesion of Ni to the original substrate. (Instead of tape, the stack could then be bonded to a silicon wafer and the original stack etched away.) Figure 4b shows a typical Raman spectrum of the GoI supported by the heat-release tape after the removal of Ni and Ni-silicide layers in HCl solution. For comparison, a Raman spectrum of the as-grown graphene from the same sample is also shown. As can be seen, the FWHM value of the GoI on tape has increased to 51.8 $cm^{-1}$ in comparison with 46.6 $cm^{-1}$ of the as-grown graphene. There is also a considerable increase of the D peak intensity, which suggests possible damage to the GoI. This is seen in Fig. 4b.1 and b.2 and could be caused by the tape peeling process as well as damage caused by the sputtering of the new insulator.

## 4. Conclusions

We have shown that large area graphene films and substrates can be created in a single-stage sputtering process. Graphene can be grown on sputtered stacks of Ni/SiC or SiC/Ni with in-situ or ex-situ rapid thermal processing at temperatures from 650 $^o$C to 1000 $^o$C. It was found that graphene always grows on the top surface of the stack, in close contact with the Ni or Ni-silicide. Raman spectra typical of high quality exfoliated monolayer graphene were obtained for samples under optimised conditions. A fast cooling rate was found to be essential to the formation of monolayer graphene. Samples with Ni atop SiC produced the best monolayer graphene spectra with ~40% surface area coverage, whereas samples with Ni below SiC produced poorer quality graphene but 99% coverage. We have also demonstrated the feasibility of transferring the grown graphene to a different insulator substrate by vacuum deposition of insulator materials directly onto the graphene surface, opening up a route for industrial scale processing of GoI wafers.

## Acknowledgment

The authors wish to acknowledge Mr Evgeny Alexeev for his assistance with Raman mapping.

# Supplementary information

## Growth of large area graphene from sputtered films

*Genhua Pan[1]\*, Mark Heath[1,2], David Horsell[2] and M. Lesley Wears[2]*

[1] Wolfson Nanomaterials and Devices Laboratory, Faculty of Science and Technology, University of Plymouth, Devon, PL4 8AA, UK

[2] College of Engineering, Mathematics and Physical Sciences, University of Exeter, Exeter, EX4 4QF, UK

### 1. Rapid thermal processing (RTP) and heating/cooling curves

After deposition of the SiC/Ni or Ni/SiC stacks, RTP was carried out either in-situ or ex-situ. Fig.S1 shows the typical heating and cooling curves for the in-situ (red curve) and ex-situ RTP using various conditions (curves of other colours).

Only one in-situ RTP condition was used in this work (due to limitations imposed by the equipment): heating the sample in vacuum at full power for 10 minutes, followed by natural cooling at zero power. As can be seen from Fig.S1, it took about 3 minutes for the temperature to reach 600 $^{o}$C. The heating rate then flattened out over the next 7 minutes, with the sample reaching a final temperature of 650 $^{o}$C. The cooling rate followed an exponential decay curve and it took more than 3 minutes for the temperature to drop down to

---

\* Corresponding author: Tel: +441752586258. Email: gpan@plymouth.ac.uk

300 °C. The temperature was measured by a thermo-couple mounted on the deposition side of the wafer surface.

The ex-situ RTP apparatus allowed for a much faster heating and cooling rate and also more control over the RTP parameters. We found that the best RTP parameters for the samples presented here were a ramp-up and ramp-down rate of 15%/s, maximum intensity of 68% (of the total power per second) and annealing time of 120s, which resulted in a typical heating cooling curve shown in Fig.S1 (solid blue curve). As can be seen, for such a condition the temperature of the sample reached a maximum of 1000 °C almost instantly and the cooling down was also much faster than in the in-situ case. Samples were also prepared using other RTP conditions, two typical ones of which are also shown in Fig.S1: one with the same heating rate, maximum intensity and annealing time, but at a slowed down cooling rate of 1%/s (dotted blue curve); another with a reduced maximum power intensity of 10% but with other parameters kept the same (brown curve). In the latter case, the heating-up time increased considerably and the maximum annealing temperature was reduced to 700 °C. Figure 3 in the main text shows a comparison of Raman spectra of samples prepared under these different RTP conditions.

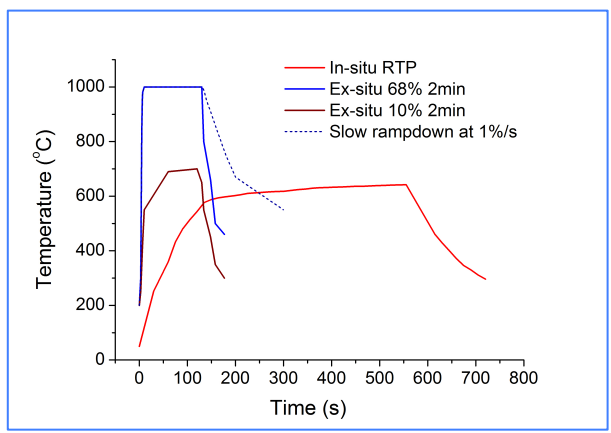

**Fig.S1** - Typical heating and cooling curves for the in-situ and ex-situ RTP.

## 2. Bragg-Brentano geometry and low angle grazing-incidence diffraction patterns of x-rays

X-ray diffraction (XRD) was carried out on selected samples with Bragg-Brentano geometry and grazing-incidence diffraction (GID) at an angle of 1 degree. The XRD measurements were conducted using a Bruker D8 Advance with a Cu tube. The GID is capable of probing the structure of very thin layers on the sample surface without picking-up the very strong diffraction signal from the Si substrate because the penetration depth of x-rays in grazing-incidence is reduced to 1~10nm from 1~10 μm in the Bragg-Brentano geometry. We studied the XRD patterns of the two different diffraction modes for a sample with clearly defined dark (graphene) and lighter (non-graphene) areas, as that shown in Fig. 2(e), but with layer structure of sub/Ni(500nm)/SiC(50nm), to see whether there is a connection between the graphene formation and crystal phase and orientation. Figure S2(a) shows the GID patterns from the dark area (blue) and lighter area (red). Although there were considerable differences in the intensities of the diffraction peaks for the two areas, it can be seen that the peak position and relative intensity of the three peaks in each diffraction pattern are very similar. This suggests that the crystal phase and their preferred orientations in the two areas are alike and there is no connection between the graphene formation and the crystal orientation. Figure S2(b) shows the XRD patterns of the two areas with Bragg-Brentano configuration diffraction, in which the very strong diffraction signals from the Si(400) is dominant due to the deep penetration depth of the x-rays into the sample. In contrast, the Ni(111) and Ni(200) peaks are hardly visible and the $Ni_{17}Si_3$(220), which was dominant in the grazing-incidence diffraction of Fig. S2(a), became invisible. This shows that for samples with SiC on top of the Ni layer, the top surface is richer in Si and there is a composition gradient due to the insufficient diffusion cross the depth of the sample.

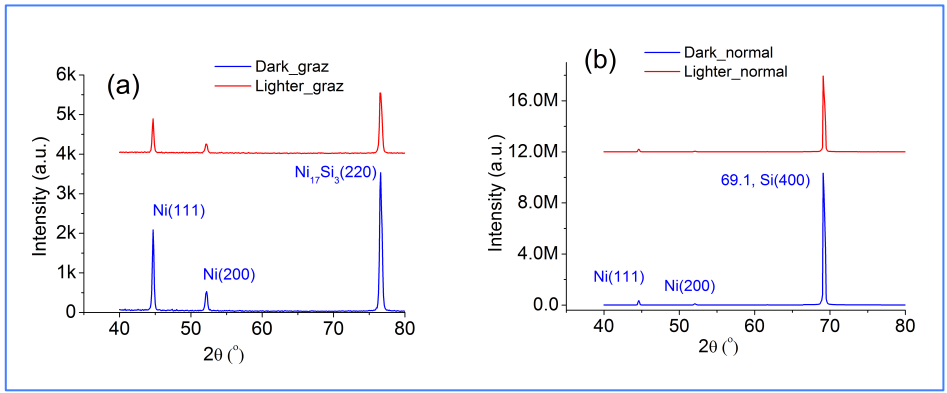

**Fig.S2 -** X-ray diffraction patterns from low angle grazing-incidence (a) and Bragg-Brentano geometry (b) of a sample with layer structure of sub/Ni(500nm)/SiC(50nm). Two areas of the sample were measured in each case: graphene (dark) and amorphous carbon (lighter).

## 3. Growth of graphene from C/Ni and SiC/Pt films

Graphene can also be produced from C/Ni and SiC/Pt films, which suggests that SiC is merely one of the carbon sources which can be employed for the growth of graphene and other metals may also be used in place of Ni for such a purpose. Here we show in Fig.S3 our initial results by using these two alternative materials. In Fig.S3, the result for the SiC(50nm)/Pt(500nm) (red) was produced by ex-situ RTP and that for C(30nm)/Ni(200nm) (blue) was produced by in-situ RTP. No further optimisation has yet been carried, but the results do indicate the presence of a graphitic layer.

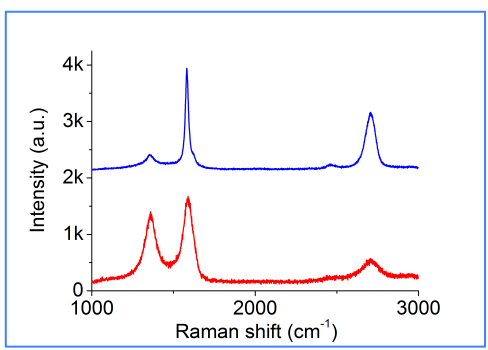

**Fig.S3** - Raman spectra for samples with layers of C(30nm)/Ni(200nm) (blue) and SiC(50nm)/Pt(500nm) (red).

## 4. Transfer of graphene to a new insulator substrate by vacuum deposition

Transfer was accomplished with the deposition of a tri-layer stack of $SiO_2$(200nm)/Ti(200nm)/$SiO_2$(200nm) on to an as-grown graphene sample prepared by ex-situ RTP. The Ti layer serves as a laser beam blocking layer for the purpose of Raman scan of the graphene-on-insulator (GoI). Ti was chosen because it does not dissolve in HCl and, at the same time, provides good adhesion to the $SiO_2$ layers. Another transfer method we investigated was to deposit 300 nm $SiO_2$ immediately after the RTP. However, after etching away the metal in HCl, the graphene/$SiO_2$ was found to roll up. Heat-release tape placed on top of the $SiO_2$ helped stabilise the sample but this prevented any Raman measurements because of the large background signal the tape introduced.

Another issue for achieving vacuum transfer of graphene in the same evacuation as the RTP is the need for a minimum number of 4 sputtering materials (Ni, SiC, $SiO_2$ and Ti) in a machine equipped with in-situ RTP. This was not possible to achieve in the 3-target machine we have without a vacuum break (change of sputtering target). For this reason, we only demonstrated the GoI process with a sample prepared by ex-situ RTP. Ultimately, the GoI needs to be prepared with in-situ RTP, in-situ deposition of $SiO_2$, direct Si wafer bonding and the removal of the original Si substrate by Si etcher, as shown in Fig.1(d) and (e).